\documentclass[11pt,a4paper]{article}
\usepackage{jcappub}

\usepackage[T1]{fontenc}

\usepackage{amsmath}
\usepackage{mathtools}
\usepackage{graphicx} 
\graphicspath{{figures/}}
\usepackage{easyReview}

\usepackage[noabbrev,nameinlink]{cleveref}

\title{\boldmath Minkowski Functionals in $SO(3)$ for the spin--2 CMB polarisation field}

\newcommand{\expec}[1]{\ensuremath{\mathbb{E}\left[#1\right]}}
\newcommand{\hess}{\ensuremath{\mathcal{H}}}
\newcommand{\eone}{\frac{\sqrt{1 - \sin{\theta}} + \sqrt{\sin{\theta} + 1}}{2\sqrt{2} \cos{\theta}} \frac{\partial}{\partial \phi } + 
    \frac{\sqrt{1 - \sin{\theta}} - \sqrt{\sin{\theta} + 1}}{2\sqrt{2} \cos{\theta}} \frac{\partial}{\partial \zeta } }
\newcommand{\etwo}{\frac{1}{\sqrt{2}} \frac{\partial }{\partial \theta }} 
\newcommand{\ethree}{\frac{\sqrt{1 - \sin{\theta}} - \sqrt{\sin{\theta} + 1}}{2\sqrt{2} \cos{\theta}} \frac{\partial}{\partial \phi } + 
    \frac{\sqrt{1 - \sin{\theta}} + \sqrt{\sin{\theta} + 1}}{2\sqrt{2} \cos{\theta}} \frac{\partial}{\partial \zeta }}
\newcommand{\eonephi}{\frac{\sqrt{1 - \sin{\theta}} + \sqrt{\sin{\theta} + 1}}{2\sqrt{2} \cos{\theta}} \frac{\partial}{\partial \phi }}
\newcommand{\eonezeta}{\frac{\sqrt{1 - \sin{\theta}} - \sqrt{\sin{\theta} + 1}}{2\sqrt{2} \cos{\theta}} \frac{\partial}{\partial \zeta }}
\newcommand{\ethreephi}{\frac{\sqrt{1 - \sin{\theta}} - \sqrt{\sin{\theta} + 1}}{2\sqrt{2} \cos{\theta}} \frac{\partial}{\partial \phi }}
\newcommand{\ethreezeta}{\frac{\sqrt{1 - \sin{\theta}} + \sqrt{\sin{\theta} + 1}}{2\sqrt{2} \cos{\theta}} \frac{\partial}{\partial \zeta }}

\author[a,b,1]{J. {Carrón Duque},\note{Corresponding author.}}
\author[a,b]{A. Carones,}
\author[c]{D. Marinucci,}
\author[a,b]{M. Migliaccio,}
\author[a,b]{and N. Vittorio}

\affiliation[a]{Dipartimento di Fisica, Università di Roma ``Tor~Vergata'', via della Ricerca Scientifica 1, I-00133, Roma, Italy}
\affiliation[b]{Sezione INFN Roma~2, via della Ricerca Scientifica 1, I-00133, Roma, Italy}
\affiliation[c]{Dipartimento di Matematica, Università di Roma Tor Vergata, via della Ricerca Scientifica 1, I-00133, Roma, Italy}

\emailAdd{javier.carron@roma2.infn.it}
\emailAdd{alessandro.carones@roma2.infn.it}
\emailAdd{marinucc@mat.uniroma2.it}
\emailAdd{marina.migliaccio@roma2.infn.it}
\emailAdd{nicola.vittorio@uniroma2.it}

\abstract{The study of the angular power spectrum of Cosmic Microwave Background (CMB) anisotropies, both in intensity and in polarisation, has led to the tightest constraints on cosmological parameters. However, this statistical quantity is not sensitive to any deviation from Gaussianity and statistical isotropy in the CMB data. Minkowski Functionals (MFs) have been adopted as one of the most powerful statistical tools to study such deviations, since they characterise the topology and geometry of the field of interest.
In this paper, we extend the application of MFs to CMB polarisation data by introducing a new formalism, where we lift the spin $2$ polarisation field to a scalar function in a higher-dimensional manifold: the group of rotations of the sphere, $SO(3)$. Such a function is defined as $f = Q \cos(2\zeta) - U \sin(2\zeta)$. We analytically obtain the expected values for the MFs of $f$ in the case of Gaussian isotropic polarisation maps. Furthermore, we present a new pipeline which estimates these MFs from input HEALPix polarisation maps. We apply it to CMB simulations in order to validate the theoretical results and the methodology. 
The pipeline is to be included in the publicly available Python package \href{https://github.com/javicarron/pynkowski}{\texttt{Pynkowski}}.}


\begin{document}
\maketitle
\flushbottom

\section{Introduction}
\label{s:intro}
The Cosmic Microwave Background (CMB) encodes information from the Early Universe, both in the intensity and polarisation of the light. The CMB polarisation field is usually decomposed into two distinct rotationally-invariant fields: the $E$ and $B$ modes \citep[see][for details]{1997PhRvL..78.2058K, 1997PhRvD..55.1830Z}. These fields are commonly studied through their angular power spectra (equivalently, $2$--point correlation functions). However, this tool is not sensitive to the possible presence of non--Gaussianities or departures from statistical isotropy of the CMB anisotropy fields.

Non--Gaussianity is predicted by many inflationary models \citep{bartolo2004, pitrou2008, barrow2006} and could shed new light on our knowledge of the primordial Universe. There is also a growing amount of literature on a possible large--scale anisotropy of the Universe, with dipoles being measured in several observables. Furthermore, the CMB maps contain foregrounds contamination because of Galactic emission and the lensing of CMB photons due to their interaction with the Large Scale Structure. These effects significantly deviate from the hypothesis of Gaussianity and isotropy, and thus have to be carefully considered when analysing the data. These effects are especially important in CMB polarisation.

Minkowski Functionals (MFs) are one of the tools adopted by the Cosmology community to study possible deviations from Gaussianity or statistical isotropy. These functionals encode geometrical and topological information of the field, not reflected in the power spectra. Other tools include the bispectrum and trispectrum, or, equivalently, the $3$-- and $4$--points correlation functions \citep{1999PhRvD..59j3001S,2004PhR...402..103B,2010JCAP...10..004B}, the distribution of maxima and minima \citep{2011PhRvD..84h3510P,carronduque2019,10.1093/mnras/stab368}, or the distribution of nonpolarised points in polarisation fields \citep{2021PhRvD.104b3502K}. MFs present several advantages with respect to the bispectrum and the trispectrum, such as the computational cost, the ease of masking or weighting data, and the possibility of studying deviations at different thresholds. The last one makes MFs naturally suited to study non--Gaussianities that are not optimally expressed in terms of momenta expansion ($f_{NL}$, $g_{NL}$, \dots); this is the case, for example, in inflationary models that can produce primordial black holes, such as Stochastic Inflation, as this introduces non--Gaussianity mostly at high values of the field \cite{figueroa2021,cruces2022}.

The application of MFs has been mostly limited to scalar maps so far, such as CMB temperature \citep{schmalzing1998,planck2019vii}, and weak lensing \cite{vicinanza2019,zorrilla2020,zuercher2021}. They have also been used to study the morphological properties of Galactic emission, like thermal dust \citep{herviascaimapo2022}, and synchrotron \citep{2021rahman,martire2023}. However, the CMB polarisation field is a complex spin $2$ quantity and MFs have not been defined for this kind of maps. In polarisation studies, MFs are usually applied to the $E$ and $B$ scalar maps independently \citep{2015JCAP...02..028G,2016JCAP...07..029S,planck2019vii}, or directly to the $Q$ and $U$ maps, ignoring spin effects \citep{2017chingangbam,puglisi2020, krachmalnicoff2020}. 

In a previous work \citep{2022arXiv221107562C}, we focused on the application of MFs to the squared polarised intensity of the CMB, $P^2=Q^2+U^2$. We introduced the formalism and computed the theoretical expectations in the Gaussian isotropic case by making use of the Gaussian Kinematic Formula. We also developed a Python package to estimate the MFs on arbitrary HEALPix scalar maps and compare them with the theoretical predictions; this software, called \texttt{Pynkowski}, is now publicly available\footnote{\href{https://github.com/javicarron/pynkowski}{https://github.com/javicarron/pynkowski}}.

In this work, we introduce a new theoretical framework to analyse the full information of the polarisation data, \textit{i.e.}, without limiting the analysis to scalar quantities defined on the sphere such as $P$, or the $E$ or $B$ modes. Such an approach provides more complete information of the statistics of CMB polarisation data, while also avoiding leakage contamination in the decomposition of masked $Q$ and $U$ maps into $E$ and $B$ modes. This is explicitly done by lifting the polarisation field to a three--dimensional space. We will therefore use MFs on three dimensions, for which there is a large body of literature \citep{gott1986,tomita1986,matsubara2003,hikage2006}. However, previous works are limited to fields defined on $\mathbb{R}^3$, while for our purposes we shall generalise this formalism for arbitrary manifolds.

The paper has the following structure. In \Cref{s:lift} we introduce the field on which the MFs will be computed and we explain some of the technical aspects needed to perform such computations. In \Cref{s:sotheo} we obtain the theoretical expectations of MFs for Gaussian and statistically isotropic spin maps. In \Cref{s:soimpl} we present the pipeline to estimate the MFs on arbitrary HEALPix spin maps from the $Q$ and $U$ data. In \Cref{s:data} we introduce the simulations %and data 
we use to validate the formalism and the pipeline, while in \Cref{s:results_so3} we present the results of applying this framework and software to the aforementioned CMB polarisation simulated maps. Finally, in \Cref{s:concl} we summarise our conclusions.

\section{Spin field as a scalar field in $SO(3)$}

\label{s:lift}
The CMB temperature anisotropies map can be seen as a real scalar field defined on the sphere. Thus, its statistical properties can be analysed with plenty of tools, such as the MFs formalism, first introduced in the Cosmological literature in \cite{schmalzing1998}. This tool is used to describe several characteristics of the excursion sets of the fields at different thresholds, which define their geometry and topology. However, the CMB polarisation has a different geometrical structure: it constitutes a complex spin $2$ field on the sphere \citep{1997PhRvD..55.1830Z}, for which excursion sets cannot be directly defined. To overcome this issue, we lift the field to a higher-dimensional space where it can be seen as a scalar field, following the framework introduced in \cite{stecconi2021}. See also \cite{lerario2022} for further mathematical discussion on spin random fields.

Let $Q(\phi, \theta)$ and $U(\phi, \theta)$ be the maps of Stokes parameters for linear polarisation in the usual base. We define $f(\phi, \theta, \zeta) : SO(3) \rightarrow \mathbb{R}$ as:
\begin{equation}
	f(\phi, \theta, \zeta) = Q(\phi, \theta) \cos(2\zeta) - U(\phi, \theta)\sin(2\zeta),
 \label{eq:fdef}
\end{equation}
which can be interpreted as the linear polarisation that one would observe at the point on the sky $(\phi, \theta)$ when observing along the polarisation direction $\zeta$, which is just an additional coordinate of the ambient space. Specifically, $f$ corresponds to the value of $Q$ if the local reference frame was rotated by an angle $\zeta$. We note that the value of $\zeta$ where $f$ is maximum for fixed $(\phi, \theta)$ is the physical polarisation angle $\psi$. The field $f$ is a three--dimensional scalar field, for which excursion sets and MFs are properly defined, as we will see in the next section.

The domain of this function, $SO(3)$, requires further technical discussion, to which we dedicate the remainder of this section. Keeping in mind the interpretation of the function variables $(\phi, \theta, \zeta)$ as the position and the polarisation direction, it can be seen that the domain of $f$ must cover all points of the sphere and all possible polarisation directions. The domain must then be contained in a three--dimensional hypersphere, $\mathbb{S}^3$. Since the CMB polarisation is a spin $2$ field, we can perform the identification $(\phi, \theta, \zeta) \cong (\phi, \theta, \zeta+\pi)$. Therefore, the domain can actually be seen as half a $3$-sphere; this space is diffeomorphic to $SO(3)$, the set of rotations of the sphere. The coordinates are the longitude $\phi\in[0,2\pi]$, the latitude $\theta\in [-\frac{\pi}{2},\frac{\pi}{2}]$, and the polarisation direction $\zeta\in[0,\pi]$. We note that any parametrisation of $SO(3)$ must present singularities. In this case, it specifically fails at $\theta=\pm\frac{\pi}{2}$ (corresponding to the poles of $\mathbb{S}^2$). However, these points constitute a zero--measure set, and therefore they do not affect the computation of the MFs, since, as discussed in the next Section, they are integrated quantities.

Following this framework, we can lift the complex spin field on the sphere to a complex scalar field on $SO(3)$. The real and imaginary parts of this field are just translations of each other, so it is enough to study only the real part in order to characterise the geometry and topology of the polarisation field. This real part is what we have called $f$ in \cref{eq:fdef}. A more detailed discussion of such lifted field can be found in \Cref{ap:f}, including the equivalence between its real and imaginary parts.

An important consequence of this construction is that if we consider an isotropic spin $2$ field on the sphere, it does not constitute an isotropic field on $\mathbb{S}^3$ nor $SO(3)$. Physically, this can be seen as a consequence of the different behaviour of the polarisation direction coordinate and the sky coordinates. In particular, the function $f$ has a deterministic (sinusoidal) dependence on the $\zeta$ variable; this will have important consequences in the predictions of the MFs of this field, as we will see in \Cref{sec:GKF}. The treatment of this anisotropy in the MF formalism represents one of the main novelties introduced in this work. Mathematically, an isotropic random field on $\mathbb{S}^3$ has to be invariant in law to the action of any isometry of $\mathbb{S}^3$; it can be proven that this is only satisfied if every multipole component of the field is a sum of fields with spin $s=-\ell, -\ell+1,\dots, \ell -1, \ell$, each with equal power. This cannot be the case for fields produced by lifting a spin field, such as CMB polarisation, since all multipoles are constituted only by $s=\pm2$. The details of this construction and the consequence on random fields can be found in \cite{stecconi2021}; some statistical properties of fields where the spin increases with the multipole can be found in \cite{lerario2022}.

Although $f$ is not isotropic in $SO(3)$, we still have enough information to produce accurate predictions for its MFs. We need simply to assume isotropy on the sphere ($\mathbb{S}^2$) and the fact that we have a spin $2$ field (\textit{i.e.}, knowledge of the behaviour of the polarisation direction coordinate). We will exploit these aspects in the next section.

To correctly compute the derivatives of the field needed to estimate the MFs, we have to take into account the geometry of the ambient manifold. The metric of $SO(3)$ in this framework is given by: 
\begin{equation}
\label{eq:metric}
	g_{\mu\nu} = 
	\begin{pmatrix}
		2 & 0 & 2\sin(\theta) \\
		0 & 2 & 0 \\
		2\sin(\theta) & 0 & 2
	\end{pmatrix},
\end{equation}
where the order is $(\phi, \theta, \zeta)$; see \Cref{ap:SOhess} for details on the computation. With this metric, the volume element and the total volume of $SO(3)$ are:
\begin{align}
	dV =& \cos(\theta)\, d\phi\,d\theta\,d\zeta ,\\ 
	\int^{2\pi}_0 \int^{\frac{\pi}{2}}_{-\frac{\pi}{2}} \int_{0}^{\pi}& \cos(\theta)\, d\phi\,d\theta\,d\zeta = 4\pi^2  .\label{eq:volso}
\end{align}

A consequence of the geometry of this manifold is that the base of the tangential space given by the derivatives $\left\{ \frac{\partial}{\partial \phi }, \frac{\partial}{\partial \theta }, \frac{\partial}{\partial \zeta }\right\}$ is not orthonormal. Therefore, it will be useful to work in the orthonormal basis $\{e_1, e_2, e_3\}$, built considering the square root of the inverse of the metric tensor:
\begin{subequations}
\begin{align}
    e_1 &= \eone ,\\
    e_2 &= \etwo ,\\ 
    e_3 &= \ethree .
\end{align}
\label{eq:sograd}
\end{subequations}

In this basis, the gradient and the Hessian of a function $f$ can be expressed as
\begin{equation}
	\nabla f =
    \begin{pmatrix}
		e_1 \\ 
		e_2  \\
		e_3
	\end{pmatrix} f,
\end{equation}
\begin{equation}
	\hess_{ij}(f) =  \nabla^2 f(e_i,e_j) = e_i e_j f - (\nabla_{e_i} e_j) f .
\label{eq:hessian}
\end{equation}

The expressions of these operators in terms of the usual spatial derivatives $\left\{ \frac{\partial}{\partial \phi }, \frac{\partial}{\partial \theta }, \frac{\partial}{\partial \zeta }\right\}$ can be found in \Cref{ap:SOhess}, together with the computations to obtain them. These results are needed to derive the MFs of arbitrary polarisation maps, as we will see in \Cref{s:soimpl}.

\section{Minkowski Functionals in $SO(3)$}
\label{s:sotheo}
MFs are statistical tools that quantify the morphology produced by a scalar function. They are higher-order statistics, meaning that their value cannot be fully predicted from any $n$--point correlation function. This renders MFs a useful complementary tool to the angular power spectrum. Although typically used on the sphere, they can be employed on any manifold, as we show in this section for the case of $SO(3)$.

\subsection{Definition}
Let $f: SO(3) \rightarrow \mathbb{R}$ be a $\mathcal{C}^2$ function, and $u\in\mathbb{R}$. The excursion set of $f$ at threshold $u$, called $A_u(f,SO(3))$, is defined as the collection of all points in the domain where the function is larger than (or equal to) $u$:
\begin{equation}
	A_{u}(f,SO(3))=\left\{ x\in SO(3):f(x)\geq u\right\}.
\end{equation} 
In the following, we will refer to it simply as $A_u$ to simplify the notation. We will use the standard notation $\partial A_u$ to refer to the boundary of $A_u$, corresponding to the points with a value of exactly $u$.

In the case of the sphere or any 2--dimensional manifold, we can define three independent MFs to fully characterise the morphology of the excursion set, as shown in \cite{schmalzing1998}. In a $3$--dimensional manifold, like $SO(3)$, we can define four independent MFs:
\begin{subequations}
\begin{align}
	V_{0}(A_{u}) &=\int_{A_{u}}dx \label{eq:V0so} ,\\ 
	V_{1}(A_{u}) &=\frac{1}{6} \int_{\partial A_{u}}da  \label{eq:V1so},\\ 
	V_{2}(A_{u}) &=\frac{1}{6\pi }\int_{\partial A_{u}}\, H(a)\,da \label{eq:V2so}, \\ 
	V_{3}(A_{u}) &=\frac{1}{4\pi }\int_{\partial A_{u}}\, K(a)\,da  \label{eq:V3so},
\end{align}
\end{subequations} 
where $da$ denotes an area element along the boundary surface $\partial A_u$, $H(a)$ and $K(a)$ are, respectively, the mean curvature (average of the two principal curvatures) and the Gaussian curvature (product of the two principal curvatures)  at a point $a$ on the boundary surface. In general, in a $n$-dimensional space, we can define $V_0$ as the hypervolume of the excursion set, and $n$ independent MFs as integrals on the boundary of a combination of the principal curvatures; see \cite{schmalzing1998} for more details. We follow the normalisation convention in this reference, noting that other normalisations have been used before, more notably in \cite{adler2007,fantaye2015}.

MFs can be interpreted as geometrical descriptors of the excursion set as a function of threshold, and therefore they represent a statistical characterisation of the original function. $V_0$ is the total volume of the excursion set. $V_1$ is the area of its boundary. To our knowledge, $V_2$ does not have a further interpretation beyond the average mean curvature of the boundary of the excursion set. Finally, $V_3$ is connected to the Euler--Poincaré characteristic $\chi$ due to an extension of the Gauss--Bonnet theorem known as the Chern--Gauss--Bonnet theorem \citep[originally proven in][]{chern1944}, which states:
\begin{equation}
	\chi(A_u) = V_3(A_u) + \frac{1}{4} V_1(A_u) \label{eq:epcso}.
\end{equation} 

Similarly to \cite{2022arXiv221107562C}, we introduce the Lipschitz--Killing curvatures, as the theoretical predictions are obtained for these quantities (see \Cref{s:theory}), which are defined as follows. Let us consider a tube of width $\rho$ built around the manifold $A_{u}$ as all the points at a distance less than $\rho$ from $A_{u}$. Its volume can be exactly expressed as a finite Taylor expansion on $\rho$, whose coefficients correspond to the Lipschitz--Killing curvatures of $A_{u}$ (see \cite{adler2007, fantaye2014} for a detailed description). These quantities are connected to the MFs in the following way:
\begin{subequations}
	\begin{align}
		\mathcal{L}_{3}(A_{u}) &= V_{0}(A_{u}) ,\\
		\mathcal{L}_{2}(A_{u}) &= 3V_{1}(A_{u}) ,\\
		\mathcal{L}_{1}(A_{u}) &= 6V_{2}(A_{u}) ,\\
		\mathcal{L}_{0}(A_{u}) &= V_3 (A_{u}).
	\end{align}
\end{subequations}

In order to compute the theoretical predictions for a Gaussian field, we have to invoke the Gaussian Kinematic Formula, as we shall see in the following section.

\subsection{Gaussian Kinematic Formula}
\label{sec:GKF}
The theoretical expected values of the Lipschitz--Killing curvatures (and therefore of MFs) for a Gaussian field can be computed with the Gaussian Kinematic Formula, as done in \cite{2022arXiv221107562C} for the polarised intensity of the CMB. 

In our case, we want to study the function $f = Q(\phi,\theta) \cos(2\zeta) - U(\phi,\theta)\sin(2\zeta)$. This function is not fully isotropic, as the behaviour in the $\theta$ and $\phi$ directions is different from the behaviour in the $\zeta$ direction. This is connected to the fact that spin $s$ functions on the sphere correspond to functions on $SO(3)$ with a single spin component $s$, but isotropic functions of $SO(3)$ must have components at all spins, and with the same angular power spectrum, as explained in \Cref{s:lift}. We shall see that this anisotropy introduces some non--negligible changes in the computations of the expected values for the MFs.

Without loss of generality, we normalise the function $f$ to have unit variance; if $Q$ and $U$ have the same variance (as required by isotropy on the sphere), this is equivalent to normalising both of them to unit variance. As a consequence of spherical isotropy, the variances of all first derivatives are equal, and the covariance between orthogonal directions is zero:
\begin{subequations}
	\begin{align}
		&\expec{\left(\frac{\partial Q}{\partial \theta}\right)^2} = \expec{\left(\frac{\partial U}{\partial \theta}\right)^2}  %= \nonumber\\
		=\expec{\left(\frac{\partial Q}{\partial \phi}\right)^2} = \expec{\left(\frac{\partial U}{\partial \phi}\right)^2} =  \mu ,\\
		&\expec{\frac{\partial Q}{\partial \theta} \cdot \frac{\partial Q}{\partial \phi}} = \expec{\frac{\partial U}{\partial \theta} \cdot \frac{\partial U}{\partial \phi}} = 0.
\end{align}
\end{subequations}
The value of $\mu$ can be directly computed as follows:
\begin{equation}
    \mu =\sum_{\ell }\frac{2\ell +1}{4\pi }\frac{(\ell-s) (\ell +s+1)}{2}C_{\ell }\text{ ,}
\label{eq:mu}
\end{equation}
where $s=2$ and:
\[
C_{\ell } = \frac{1}{2}\Big(C^{EE}_{\ell }+C^{BB}_{\ell }\Big),
\]
with $C^{EE}_{\ell }$ and $C^{BB}_{\ell }$ the EE and BB angular power spectra, respectively, computed from Q and U maps. We note that $\mu\gg1$, as a direct consequence of normalising the field to unit variance.

The last quantity we need to introduce is the covariance matrix of the derivatives of $f$, which can be defined as ${\Sigma}_{ij} = \expec{\frac{\partial f}{\partial e_i} \cdot \frac{\partial f}{\partial e_j}}$, using the orthonormal basis $\{e_1, e_2, e_3\}$, introduced in \cref{eq:sograd}. Specifically, its determinant is:
\begin{equation}
    |\Sigma|^{\frac{1}{2}} = \frac{5}{2\sqrt{2}}\,\mu .
\end{equation}
We note that, in a fully isotropic case, $|\Sigma|^{\frac{1}{2}}$ would scale as $\mu^{3/2}$. Thus, the anisotropy of $f$ will introduce different $\mu$--scaling factors in the prediction of MFs. This anisotropy is associated to the different behaviour of the field $f$ in the $(\theta,\phi)$ coordinates (random Gaussian) and in the $\zeta$ coordinate (deterministic sinusoidal).

At this point, the Gaussian Kinematic Formula can be formulated for these non--isotropic Gaussian functions with a caveat: the Lipschitz--Killing curvatures must be computed not with the usual metric of the ambient manifold (see \cite{schmalzing1998,2022arXiv221107562C}), but with a metric given by the covariance matrix $\Sigma$ introduced above. We denote $\mathcal{L}_{i}^f(A_{u})$ the Lipschitz--Killing curvatures computed with this metric. For the Gaussian function $f:M\rightarrow\mathbb{R}$, the Gaussian Kinematic Formula takes the following form:
\begin{equation}
	\expec{ \mathcal{L}_{i}^f (A_{u}) } = \sum_{k=0}^{dim(M)-i}%
	\begin{bmatrix}
		k+i \\ 
		k
	\end{bmatrix}%
	\rho_{k}(u)\mathcal{L}_{k+i}^f (M)  ,\label{eq:gkfso}
\end{equation}%
where the flag coefficients are \citep{adler2007}:%
\[
\left[
\begin{array}{c}
k+j \\
k%
\end{array}%
\right] =\frac{\omega _{k+j}}{\omega _{k}\omega _{j}}\left(
\begin{array}{c}
k+j \\
k%
\end{array}%
\right) \text{ , }
\omega _{j}=\frac{\pi ^{j/2}}{\Gamma (\frac{j}{2}+1)}\text{
,}
\]%
with $\omega_j$
representing the volume of the $j$--dimensional unit ball,
while the functions $\rho_k(u)$ are defined as: 
\begin{align*}
\rho _{k}(u) =\frac{1}{(2\pi )^{k/2}}&\frac{1}{\sqrt{2\pi }}\exp \Bigg(-\frac{%
u^{2}}{2}\Bigg)H_{k-1}(u)\text{ ,} \\
H_{-1}(u) &=\sqrt{2\pi}\exp\left(\frac{%
u^{2}}{2}\right)\left(1-\Phi (u)\right)\text{ , }\\H_{0}(u)&=1\text{ , }\\H_{1}(u)&=u\text{ ,}\\H_{2}(u)&=u^2 -1\text{ ,}\\
H_{k}(u)=(-1)^{k}&\exp \Bigg(\frac{u^{2}}{2}\Bigg)\frac{d^{k}}{du^{k}}\exp \Bigg(-%
\frac{u^{2}}{2}\Bigg)\text{ .}
\end{align*}%
The function $\Phi$ represents the cumulative normal distribution and $H_k$ are the Hermite polynomials.

In the case of an isotropic function with the normalisation described above, it can be proven that $\mathcal{L}_{j}^f (M) = \mathcal{L}_{j} (M) \mu^{j/2}$, with $0\leq j\leq dim(M)$. This identity simplifies \cref{eq:gkfso} to the Gaussian Kinematic Formula described in \cite{2022arXiv221107562C}. 

In the non--isotropic case, the relation between the Lipschitz--Killing curvatures computed with both metrics is not trivial, but there are two important observations. First, $\mathcal{L}_{0}$ corresponds to the Euler--Poincaré characteristic, which is a topological invariant. This means that this quantity depends only on the topological structure and not on the metric. Therefore, $\mathcal{L}_{0}^f (M) = \mathcal{L}_{0} (M)$. Second, there is a useful identity for $\mathcal{L}_{dim(M)}(M)$: this quantity corresponds to the total volume of the manifold $M$ or, more technically, to its Hausdorff measure. Computing it with a different metric is analogous to a change of variable in the integral of the measure. Therefore, $\mathcal{L}_{dim(M)}^f (M) = |\Sigma|^{1/2} \mathcal{L}_{dim(M)} (M)$, which, thus, in our case can be rephrased by: $\mathcal{L}_{3}^f (M) = \left(\frac{5}{2\sqrt{2}}\,\mu\right)\mathcal{L}_{3} (M)$.

Note that if we ignore the anisotropy of $f$, we would obtain that the factor is $\propto \mu^{3/2}$ instead of $\propto\mu$, yielding an incorrect behaviour with $\mu$ and therefore with the angular power spectrum of the polarisation maps. This is one of the most remarkable advantages of adopting this general formalism.

Finally, the cases of $\mathcal{L}_{1}^f (M)$ and $\mathcal{L}_{2}^f (M)$ are more complicated, as they are related to the change of integrating submanifolds with lower dimensionality when the global (non-isotropic) metric changes. Our conjecture is that the scaling relations should be $\mathcal{L}_{1}^f (M) \propto \mathcal{L}_{1} (M)$ and $\mathcal{L}_{2}^f (M) \propto \mathcal{L}_{2} (M) \mu^{1/2}$. In \Cref{ss:scaling} we will see that these guesses are highly compatible with simulations, and we will compute the proportionality factors.

\subsection{Theoretical predictions}
\label{s:theory}
We are now in the position to explicitly compute the theoretical predictions for the MFs of the excursion set. In our case, the global manifold is $M=SO(3)$. We note that the right hand side of the Gaussian Kinematic Formula, \cref{eq:gkfso}, always contains the term proportional to $\mathcal{L}_{3}(SO(3))$ as the leading term. As mentioned above, the last Lipschitz--Killing curvature is equal to the Hausdorff measure (or volume) of the manifold, which in this case is $ \textrm{Vol}(SO(3))=4\pi^2$, as shown in \cref{eq:volso}. 

Only $\mathcal{L}_{3}$ is needed to compute all MFs at leading order but we report the rest of the Lipschitz--Killing curvatures of $SO(3)$ for completeness:
\begin{subequations}
	\begin{align}
			\mathcal{L}_{3}(SO(3)) &= 4\pi^2 ,\\
			\mathcal{L}_{2}(SO(3)) &= 0 ,\\
			\mathcal{L}_{1}(SO(3)) &= 6\pi              ,\\
			\mathcal{L}_{0}(SO(3)) &= 0.
		\end{align} \label{eq:lsso}
\end{subequations}% 
For the topology--inclined readers, the last line implies that the Euler--Poincaré characteristic vanishes: $\chi(SO(3))=0$; this is a consequence of the $3$-sphere being a ($2$-fold) cover of $SO(3)$ and $\chi(\mathbb{S}^3) = 0$.
 
We can now compute the Lipschitz--Killing curvatures of the excursion set of $f$ using the Gaussian Kinematic Formula, \cref{eq:gkfso}. We assume no mask in these computations, but they can be readily introduced by modifying the Lipschitz--Killing curvatures of the global manifold, most notably by multiplying the total volume by the sky fraction.

\paragraph*{Volume of the excursion set, $V_{0}$.} We use the Gaussian Kinematic Formula with $j=3$ to compute the expected value of the volume of the excursion set, along with the expression $\mathcal{L}_{3}^f (M) = \left(\frac{5}{2\sqrt{2}}\,\mu\right)\mathcal{L}_{3} (M)$ derived above: 
\begin{equation}
\begin{split}
	\expec{ \mathcal{L}_{3} (A_{u}) }& = 
		\frac{2\sqrt{2}}{5\mu} \expec{ \mathcal{L}_{3}^f (A_{u}) }\\
		&=\frac{2\sqrt{2}}{5\mu} \sum_{k=0}^{3-3}%
		\begin{bmatrix}
			k+3 \\ 
			k
		\end{bmatrix}%
		\rho_{k}(u)\mathcal{L}_{k+3}^f (SO(3)) \\
		&= \frac{2\sqrt{2}}{5\mu} %
		\begin{bmatrix}
			3 \\ 
			0
		\end{bmatrix}%
		\rho_{0}(u) \left[\frac{5}{2\sqrt{2}}\,\mu\mathcal{L}_{3} (SO(3))\right] =  4\pi^2 \left[ 1 -\Phi(u) \right],
\end{split}% 
\label{eq:l3so}
\end{equation}% 

where $\Phi(u)$ is the cumulative distribution function of the standard normal distribution. It can be seen that $V_0$ is equal to the total volume at low thresholds (since the field at all points is greater than $u$), and it is $0$ at high thresholds (since the field in all points is lower than $u$). Interestingly, the definition of $f$ implies that at $u=0$, $\mathcal{L}_{3} (A_{0}) = 2\pi^2$ with no scattering whatsoever. We note that the factor $\mu$ cancels out, and the theoretical expectation depends only on the threshold $u$, not on the field itself (as long as it is normalised to have unit variance). The summation in \cref{eq:l3so} has exactly one term, corresponding to the volume of $SO(3)$, so no approximation is needed for this prediction.

\paragraph*{Area of the boundary of the excursion set, $V_{1}$.} We can compute the theoretical expectation by setting $j=2$ in the Gaussian Kinematic Formula and assuming the scaling relation explained above: 
\begin{equation}
  \mathcal{L}_{2}^f (M) = K_1^{-1} \mu^{1/2} \mathcal{L}_{2} (M)    ,
\end{equation} 
where we have introduced the unknown constant $K_1$ (as it is related to $V_1$). Thus: 
\begin{equation}
\begin{split}
	\expec{ \mathcal{L}_{2} (A_{u}) } &= 
		K_1 \mu^{-1/2} \expec{ \mathcal{L}_{2}^f (A_{u}) } \\
		&= K_1 \mu^{-1/2} \sum_{k=0}^{3-2}%
		\begin{bmatrix}
			k+2 \\ 
			k
		\end{bmatrix}%
		\rho_{k}(u)\mathcal{L}_{k+2}^f (SO(3)) \\
		&= K_1 \mu^{-1/2}   %
		\begin{bmatrix}
			3 \\ 
			1
		\end{bmatrix}%
		\rho_{1}(u) \left[\frac{5}{2\sqrt{2}}\,\mu\mathcal{L}_{3} (SO(3))\right] \\
		&= K_1 \frac{5}{\sqrt{2}}\sqrt{\mu} \, 4\pi^2 \left[ \frac{1}{2\pi}\exp{\left(\frac{-u^2}{2}\right)} \right],
\end{split}% 
\label{eq:l2so}
\end{equation}%
where the square bracket in the fourth line corresponds to $\rho_{1}(u)$ and the term $k=0$ in the summation is zero because of \cref{eq:lsso}.

\paragraph*{Mean curvature of the boundary of the excursion set, $V_{2}$.} In a similar way, we can compute the theoretical expectation of $V_2$ by setting $j=1$ in the Gaussian Kinematic Formula and assuming the scaling relation explained above: 
\begin{equation}
    \mathcal{L}_{1}^f (M) = K_2^{-1} \mathcal{L}_{1} (M),
\end{equation} 
where we again introduce an unknown constant $K_2$. Thus:
\begin{equation}
\begin{split}
	\expec{ \mathcal{L}_{1} (A_{u}) } &= 
		K_2 \expec{ \mathcal{L}_{1}^f (A_{u}) } \\
		&= K_2 \sum_{k=0}^{3-1}%
		\begin{bmatrix}
			k+1 \\ 
			k
		\end{bmatrix}%
		\rho_{k}(u)\mathcal{L}_{k+1}^f (SO(3)) \\
		&= K_2   %
		\begin{bmatrix}
			3 \\ 
			2
		\end{bmatrix}%
		\rho_{2}(u) \left[\frac{5}{2\sqrt{2}}\,\mu\mathcal{L}_{3} (SO(3))\right] +\mathcal{O}(\mathcal{L}_{1}^f (SO(3)))\\
		&= K_2 \frac{5}{\sqrt{2}}\mu\, 4\pi^2 \left[ \frac{u}{(2\pi)^{3/2}}\exp{\left(\frac{-u^2}{2}\right)} \right] +\mathcal{O}(\mu^0),
\end{split}% 
\label{eq:l1so}
\end{equation}%
where the square bracket in the fourth line corresponds to $\rho_{2}(u)$. Only two terms in the summation are not zero, one of order $\mu$ and another of order $1$. Given that $\mu\gg1$ (in the case of cosmological fields, we typically have $\mu\sim10^5$), we neglect the second order terms in the computations.

\paragraph*{Euler--Poincaré characteristic of the excursion set, $V_{3}$.} Finally, we compute the theoretical prediction for the Euler--Poincaré characteristic of the excursion sets of $f$. As explained before, $\mathcal{L}_{0}^f (M) = \mathcal{L}_{0} (M)$, since it is a topological invariant. Thus:
\begin{equation}
    \begin{split}
        \expec{ \mathcal{L}_{0} (A_{u}) } &= 
        \expec{ \mathcal{L}_{0}^f (A_{u}) } \\
        &=\sum_{k=0}^{3-0}%
        \begin{bmatrix}
            k+0 \\ 
            k
        \end{bmatrix}%
        \rho_{k}(u)\mathcal{L}_{k+0}^f (SO(3)) \\
        &= %
        \begin{bmatrix}
            3 \\ 
            3
        \end{bmatrix}%
        \rho_{3}(u) \left[\frac{5}{2\sqrt{2}}\,\mu\mathcal{L}_{3} (SO(3))\right] %
        + \begin{bmatrix}
            1 \\ 
            1
        \end{bmatrix}%
        \rho_{1}(u) \left[\mathcal{L}_{1}^f (SO(3))\right] \\
        &=  \frac{5}{2\sqrt{2}}\,\mu\,4\pi^2 \left[ \frac{(u^2-1)}{(2\pi)^2}\exp{\left(-\frac{u^2}{2}\right)} \right] +\mathcal{O}(\mu^0),
        %&= 2\mu \, (u^2-1) \exp{\left(-\frac{u^2}{2}\right)} +\mathcal{O}(\mu^0)
    \end{split}  
 \label{eq:l0so}
\end{equation}% 
where the square bracket in the fourth line corresponds to $\rho_{3}(u)$. Again, only two terms in the summation are not zero, one of order $\mu$ and another of order $1$. The latter, the term corresponding to $\mathcal{L}_{1}^f (SO(3))$, can be safely ignored.

We note that neglecting the anisotropy of $f$ would yield an incorrect exponent for $\mu$ in the theoretical prediction for $\mathcal{L}_0$ (\textit{i.e.}, the Euler-Poincaré characteristic).

\paragraph*{Predictions for the MFs} We can now convert the predictions for Lipschitz--Killing curvatures into MFs. Additionally, in order to ease the comparison with data and the interpretation, we work with normalised MFs, where the quantities are divided over the volume (in our case this is $4\pi^2$, possibly multiplied by the sky fraction $f_{sky}$ if we impose a mask):
\begin{equation}
    v_i = \frac{V_i}{4\pi^2}.
\end{equation}

The theoretical predictions of the normalised MFs are as follows:
\begin{subequations}
\label{eq:mf_theo}
	\begin{align}
		\expec{ v_0 } &= 1 -\Phi(u) ,\\
		\expec{ v_1 } &= K_1 \, \frac{5}{6\pi\sqrt{2}} \, \mu^{1/2} \, \exp{\left(\frac{-u^2}{2}\right)},\\
		\expec{ v_2 } &= K_2 \, \frac{5}{24{\pi^{3/2}}} \, \mu\, u \exp{\left(\frac{-u^2}{2}\right)} ,\\
		\expec{ v_3 } &= \frac{5}{8\sqrt{2}\, \pi^2} \, \mu\, (u^2-1) \exp{\left(\frac{-u^2}{2}\right)}.
	\end{align}
\end{subequations}
We recall that the first two formulae are exact, while the last two are correct at the leading order in $\mu$. One of the main advantages of this formalism is that the expected values of the normalised MFs do not depend on the use of masks. These theoretical predictions show the power of this approach. We are able to predict a topological feature of the field knowing only the parameter $\mu$, which can be computed from the angular power spectra of the E and B modes. Alternatively, this parameter can also be obtained directly on the maps, which may be recommended when they are masked to avoid leakage effects, which distort the estimation of the polarisation angular power spectra. 

Similar results to \cref{eq:mf_theo} are known in the literature for both Gaussian and weakly non--Gaussian isotropic fields defined on $\mathbb{R}^3$ \citep{matsubara2003,hikage2006}. The main differences in our approach are the following:
\begin{itemize}
    \item The use of the Gaussian Kinematic Formula, which unifies the theoretical computations for all MFs.
    \item We consider a space with a non--trivial metric given in \cref{eq:metric}; this approach can be easily generalised to arbitrary manifolds. We note that this only affects the normalisation factors in \cref{eq:mf_theo}.
    \item The studied field $f$ given in \cref{eq:fdef} is not isotropic, leading to additional complications. We showed how to deal with them when the anisotropic behaviour is known. In our case, this implied a distinct trend with $\mu$ for the Euler--Poincarè characteristic.
\end{itemize}

In this work, we present only the predictions for Gaussian fields. The extension to weakly non--Gaussian fields could be obtained by adopting an approach similar to that presented in \citep{matsubara2003}, by considering higher--order cumulants of the probability density function.

\section{Implementation: \texttt{Pynkowski}}
\label{s:soimpl}
We implement both the theoretical predictions (presented in \Cref{s:sotheo}) and the computation of the MFs of $f$ on data (to be explained in this section). Both aspects are to be included in the publicly available Python package called \texttt{Pynkowski}, which can be found in \url{https://github.com/javicarron/pynkowski} and was first introduced in \cite{2022arXiv221107562C} in the context of MFs for scalar maps, such as $T$ and $P^2 = {Q^2 + U^2}$.

We develop a software to compute all the necessary quantities related to $f$, including its derivatives in all directions, the covariant gradient, and the Hessian. We use this code on Gaussian CMB simulations in order to compare the values of their MFs with the theoretical predictions. After this validation, it can be applied to arbitrary maps to assess any deviation from statistical isotropy and Gaussianity. The actual implementation of the computation of these MFs on maps is described in this section.

For a map $f$, we store the values of $Q$ and $U$ in the HEALPix pixelisation scheme \citep{gorski2005}. The variable $\zeta$ is not pixelised, it is calculated exactly for each pixel $\xi_i$ as $f(\xi_i, \zeta) = Q(\xi_i) \cos(2\zeta) - U(\xi_i)\sin(2\zeta)$. 

The first and second spatial derivatives of $Q$ and $U$ with respect to $\theta$ and $\phi$ are also obtained when needed, with the help of the healpy\footnote{\url{https://github.com/healpy/healpy}} function \texttt{alm2map\_der}, which performs such computation in harmonic space. The first and second spatial derivatives of $f$ with respect to its three variables are simplified analytically and expressed as exact functions of the polarisation maps and their derivatives. This means that the only pixelisation is due to the original one of $Q$ and $U$ and their derivatives, while the treatment in $\zeta$ is always exact in order to avoid numerical artefacts.

The first step in our procedure after loading the map is to normalise $f$ to have unit variance. We do not assume that $Q$ and $U$ have similar statistical properties, but if they do, this step is equivalent to normalising both of them to have unit variance. With such a normalisation of $f$, we can express the MFs as a function of an adimensional threshold $u$. For convenience, we compute the MFs divided by the total volume of $SO(3)$ to have the normalised MFs ($v_i = \frac{V_i}{4\pi^2}$).

Before explaining the specific computations for each MF, we shall introduce some common notation. Let $P(\xi_i) = \sqrt{Q(\xi_i)^2+U(\xi_i)^2}$, \textit{i.e.}, the maximum value of $f$ in each pixel $\xi_i$. For a fixed $\xi_i$, we note that $f$ has a simple sinusoidal behaviour in its variable $\zeta$: it is always below $u$ if $u>P(\xi_i)$, always above $u$ if $u<-P(\xi_i)$, and above $u$ in a single segment of $\zeta$ otherwise. We use this observation to split all the integrals involved in the computation of the MFs. In the case where $f(\xi_i) > u$ only in a range of $\zeta$, let $\zeta_M(\xi_i)$ be the angle at which $f(\xi_i, \zeta)$ is maximum, while $\zeta_1(\xi_i)$ and $\zeta_2(\xi_i)$ the angles for which $f(\xi_i)=u$; they can be calculated as:
\begin{subequations}
	\begin{align}
		\zeta_M(\xi_i) &= \frac{1}{2}\arctan\left(-\frac{U(\xi_i)}{Q(\xi_i)}\right) ,\\
		\zeta_1(\xi_i) &= \zeta_M(\xi_i) - \frac{1}{2} \arccos\left(\frac{u}{P(\xi_i)}\right) ,\\
		\zeta_2(\xi_i) &= \zeta_M(\xi_i) + \frac{1}{2} \arccos\left(\frac{u}{P(\xi_i)}\right) ,
	\end{align}
\end{subequations}
where the $\arctan$ in the first line is defined to be in the quadrant corresponding to the coordinate $(Q(\xi_i), U(\xi_i))$. Polarisation directions $\zeta$ and their differences are always defined between $0$ and $\pi$, due to the geometry of $SO(3)$.

\paragraph*{First MF, $\mathbf{v_0}$}
It can be seen as the volume fraction of the manifold for which $f>u$:
\begin{equation}
	v_0(u) = \frac{1}{4\pi^2} \int_{SO(3)} \Theta(f(x)-u) \, dx \label{eq:v0so},
\end{equation} 
where $\Theta(r)$ is the Heaviside function ($1$ where $r\geq0$; $0$ otherwise). We can split the integrand by pixel according to the cases explained above and integrate with respect to $\zeta$ first. The integrand will be $0$ for pixels where $f$ is always smaller than the threshold ($u>P(\xi_i)$) and it will be $1$ if $u<-P(\xi_i)$. For all the other pixels, it will be $1$ as long as $\zeta$ is between $\zeta_1$ and $\zeta_2$ and $0$ otherwise, so the integral in $\zeta$ will yield the length of the interval between these two angles, \textit{i.e.}, $\arccos\left(\frac{u}{P(\xi_i)}\right) \frac{1}{\pi}$.

Therefore, to compute \cref{eq:v0so}, we just need the length of this interval for every pixel and average it over all pixels.

\paragraph*{Second MF, $\mathbf{v_1}$} It can be computed as the area of the boundary of the excursion sets, \textit{i.e.}, the manifold defined by $f=u$:
\begin{align}
	\begin{split}
		v_1(u) &=\frac{1}{4\pi^2} \frac{1}{6} \int_{\partial A_{u}}da \\
		&=  \frac{1}{4\pi^2} \frac{1}{6}  \int_{SO(3)} \delta(f(x)-u) \cdot |\nabla f(x)| \, dx,
	\end{split}
\end{align} 
where the second equality comes from a change of coordinates from the surface element of the boundary $da$ to a volume element on $SO(3)$, $dx$. The gradient of $f$ is denoted by $\nabla f$ and can be computed with \cref{eq:sograd}; $\delta$ is the Dirac delta.

In a fully pixelised field (such as $T$ or $P$), one typically has to bin the threshold and approximate the delta function (see \cite{2022arXiv221107562C} for a discussion in that case). However, in the present framework we can exactly determine the points where $f(\xi_i, \zeta)=u$ for each pixel (if they exist) as $(\xi_i, \zeta_1)$ and $(\xi_i, \zeta_2)$.

To compute the integral, it suffices to evaluate $|\nabla f(x)|$ at these points (adding the values corresponding to both $\zeta_1$ and $\zeta_2$ for each $\xi_i$), fill with $0$ the rest of the pixels, and compute the average value of this quantity across the entire map. This result is then divided by $6$, the normalisation factor.

\paragraph*{Third MF, $\mathbf{v_2}$} It is the average mean curvature ($H$) of the boundary of the excursion sets:
\begin{align}
	\begin{split}
		v_2(u) &=\frac{1}{4\pi^2} \frac{1}{6\pi} \int_{\partial A_{u}}  H(a) da \\
		&=  \frac{1}{4\pi^2} \frac{1}{6\pi}  \int_{SO(3)} \delta(f(x)-u) \cdot H(x) \cdot |\nabla f(x)| \, dx,
	\end{split}
\end{align} 
where we use the same change of coordinates as in the previous case, which introduces the $\delta$ and the factor $|\nabla f(x)|$. 

In general, evaluating the mean curvature of a surface is far from trivial. In this case, however, the surface is defined implicitly by the function $F(x) \coloneqq f(x)-u = 0$, so we can use the expression for the mean curvature in an implicitly defined surface \citep{docarmo1976}:
\begin{equation}
	H = \frac{ \nabla F\  \hess(F) \ \nabla F^{\mathsf {T}} - |\nabla F|^2\, \text{Tr}(\hess(F)) } { 2|\nabla F|^3 } ,
\end{equation} 
where all quantities are functions of $x$.

As before, we compute the integral by evaluating $H(x) \cdot |\nabla f(x)|$ in the points where $f=u$ (adding for each pixel the values corresponding to both $\zeta_1$ and $\zeta_2$), we fill with $0$ the rest of the pixels, and calculate the average value of this quantity across the entire map. This result is then divided by $6\pi$, the normalisation factor.

\paragraph*{Fourth MF, $\mathbf{v_3}$} It can be computed as the average Gaussian curvature ($K$) of the boundary of the excursion sets:
\begin{align}
	\begin{split}
		v_3(u) &=\frac{1}{4\pi^2} \frac{1}{4\pi} \int_{\partial A_{u}}  K(a) da \\
		&=  \frac{1}{4\pi^2} \frac{1}{4\pi}  \int_{SO(3)} \delta(f(x)-u) \cdot K(x) \cdot |\nabla f(x)| \, dx ,
	\end{split}
\end{align} 
where we use the same strategy as in the previous case, now computing the Gaussian curvature $K(x)$ instead of the mean curvature $H(x)$. In the case of an implicitly defined surface (again $F(x) \coloneqq f(x)-u = 0$), the Gaussian curvature can be computed as \citep{docarmo1976}:
\begin{equation}
	K=-\frac{ 
		\begin{vmatrix}
			H(F) & \nabla F^{\mathsf T} \\
			\nabla F & 0 
		\end{vmatrix}
	}{ |\nabla F|^4 }
	=-\frac{ 
		\begin{vmatrix}
			F_{xx} & F_{xy} & F_{xz} & F_x \\
			F_{xy} & F_{yy} & F_{yz} & F_y \\
			F_{xz} & F_{yz} & F_{zz} & F_z \\
			F_{x} & F_{y} & F_{z} & 0 \\
		\end{vmatrix}
	}{ |\nabla F|^4 }.
\end{equation} 

We then compute the integral by evaluating $K(x) \cdot |\nabla f(x)|$ at the points where $f=u$ (adding for each pixel the values corresponding to both $\zeta_1$ and $\zeta_2$), we fill with $0$ the rest of the pixels, and calculate the average value of this quantity across the entire map. This result is then divided by $4\pi$, the normalisation factor.

\subsection{General considerations} 
All the above computations are performed without pixelising the variable $\zeta$, only $\theta$ and $\phi$. However, numerical issues can arise, especially in $v_3$ and, to a lesser extent, $v_2$, due to the computation of the second spatial derivatives in pixelised maps. In order to avoid artefacts, we must work with smooth maps, where the value of the field does not change abruptly between neighbouring pixels; in other words, the derivatives of the map must exist and be reasonably smooth. This can be seen as a requirement on the maps $Q$ and $U$ to have negligible power at high multipoles or on the needed pixel resolution (the $N_{\textrm{side}}$ parameter in HEALPix maps). This is not only a computational limitation but, in fact, a mathematical requirement. Indeed, the Gaussian Kinematic Formula holds if the boundaries of the excursion sets are twice differentiable (see \cite{adler2007}). In practice, this means that the angular power spectrum must decay fast enough. For a study where this condition does not hold, see, \textit{e.g.}, \cite{lan2018}.

Nevertheless, we note that the theoretical predictions for $v_2$ and $v_3$ are correct up to leading order in $\mu$. Given that $\mu$ increases with the considered range of multipoles, these approximations are increasingly accurate at higher multipoles. Therefore, it may be convenient to apply this formalism on needlet components. Indeed, needlets naturally filter out the low and high multipoles, thus reducing both the pixelisation effects and the error in the approximation in $\mu$. MFs in needlet domain were studied for scalar fields on the sphere in, \textit{e.g.}, \cite{fantaye2015,planck2019vii}.

\section{Simulations}
\label{s:data}
In this section we introduce the different polarisation fields that we analyse with MFs.

\subsection{Monochromatic maps}
\label{ss:mono}
The exact value of $v_1$ and $v_2$, introduced in \Cref{s:theory}, depends on the unknown normalisation factors $K_1$ and $K_2$, respectively.

We generate simulated maps, using the \texttt{healpy} Python package \citep{Zonca2019}, in order to verify the predicted dependence with $\mu$ of the four MFs, and compute the two unknown normalisation factors. We create $1600$ monochromatic maps, \textit{i.e.}, maps with power in a single multipole $\ell_m$:

\begin{equation}
    C_{\ell}^{EE} = C_{\ell}^{BB} = \begin{cases} 1 & \ell = \ell_{m},\\ 0 & {\rm otherwise} .\end{cases}
%\label{eq:cl_mono}
\end{equation}
We simulate $8$ sets of $200$ maps, each set with a different $\ell_m$. We have considered $\ell_m=\{ 100, 485, 675, 825, 950, \\
1060, 1165, 1255 \}$. These values of $\ell_m$ are chosen in order to have a set of approximately equally--spaced values of $\mu_m=(\ell_m-s)(\ell_m+s+1)/2$, where $s=2$.

\subsection{Realistic angular power spectrum}
\label{ss:simsgau}
In order to validate the theoretical predictions and the implemented computation of MFs on realistic maps, we generate $300$ Gaussian isotropic CMB simulations with the best fit angular power spectrum for CMB polarisation (E and B modes) reported by \cite{planck_spectra}. We use a map resolution of $N_{side}=1024$. We smooth the maps with a Gaussian beam of FWHM$=15'$ to avoid pixelisation effects in the computation of the spatial derivatives.

In this way, we are able to simultaneously achieve three goals: assess the accuracy of the theoretical formulae obtained in \Cref{s:theory}, verify the implementation of MFs computation on simulated maps, and check that the normalisation constants extend correctly to non--monochromatic maps.

\section{Results}
\label{s:results_so3}

\subsection{Normalisation constants}
In this section we determine the normalisation constants for $v_1$ and $v_2$, \textit{i.e.}, $K_1$ and $K_2$. As mentioned in \Cref{s:theory}, the exact values for the predicted MFs can only be computed for $v_0$ and $v_3$, while the prediction for $v_1$ and $v_2$ is correct except for a global normalisation factor related to the anisotropic $\zeta$ direction, see \cref{eq:mf_theo}. We note that both $K_1$ and $K_2$ are truly constants and therefore do not depend on the threshold, mask, or value of $\mu$.

We estimate the values of these constants by computing the MFs on the monochromatic maps introduced in \Cref{ss:mono} with the software presented in \Cref{s:soimpl} and soon to be included in \texttt{Pynkowski}. Then, $K_1$ and $K_2$ are obtained as the ratio between the computation on the maps and theoretical predictions, averaging over all simulations and all thresholds.
We find that the constants are (average and standard deviation):
\begin{subequations}
    \label{eq:ks}
    \begin{align}
        K_1 &= 0.31912 \pm 0.00001, \\
        K_2 &= 0.7088 \pm 0.0002.
    \end{align}
\end{subequations}

We have verified that these values do not significantly vary with the threshold or value of $\mu$ (\textit{i.e.}, multipole). We have also checked that the constants are valid for non--monochromatic maps, as we will see in \Cref{ss:gau_res}.

Lastly, we note that the theoretical predictions yield the correct normalisation factor for $v_0$ and $v_3$, as expected. If we repeat the procedure with additional multiplicative factors, $K_{v_0}$ and $K_{v_3}$, for these MFs, we see that both are compatible with $1$:
\begin{subequations}
    \label{eq:ks_1}
    \begin{align}
        K_{v_0} &= 1.0000 \pm 0.0001, \\
        K_{v_3} &= 1.001 \pm 0.003.
    \end{align}
\end{subequations}

\subsection{Scaling relations}
\label{ss:scaling}

\begin{figure}
	\centering
	\includegraphics[width=0.6\textwidth]{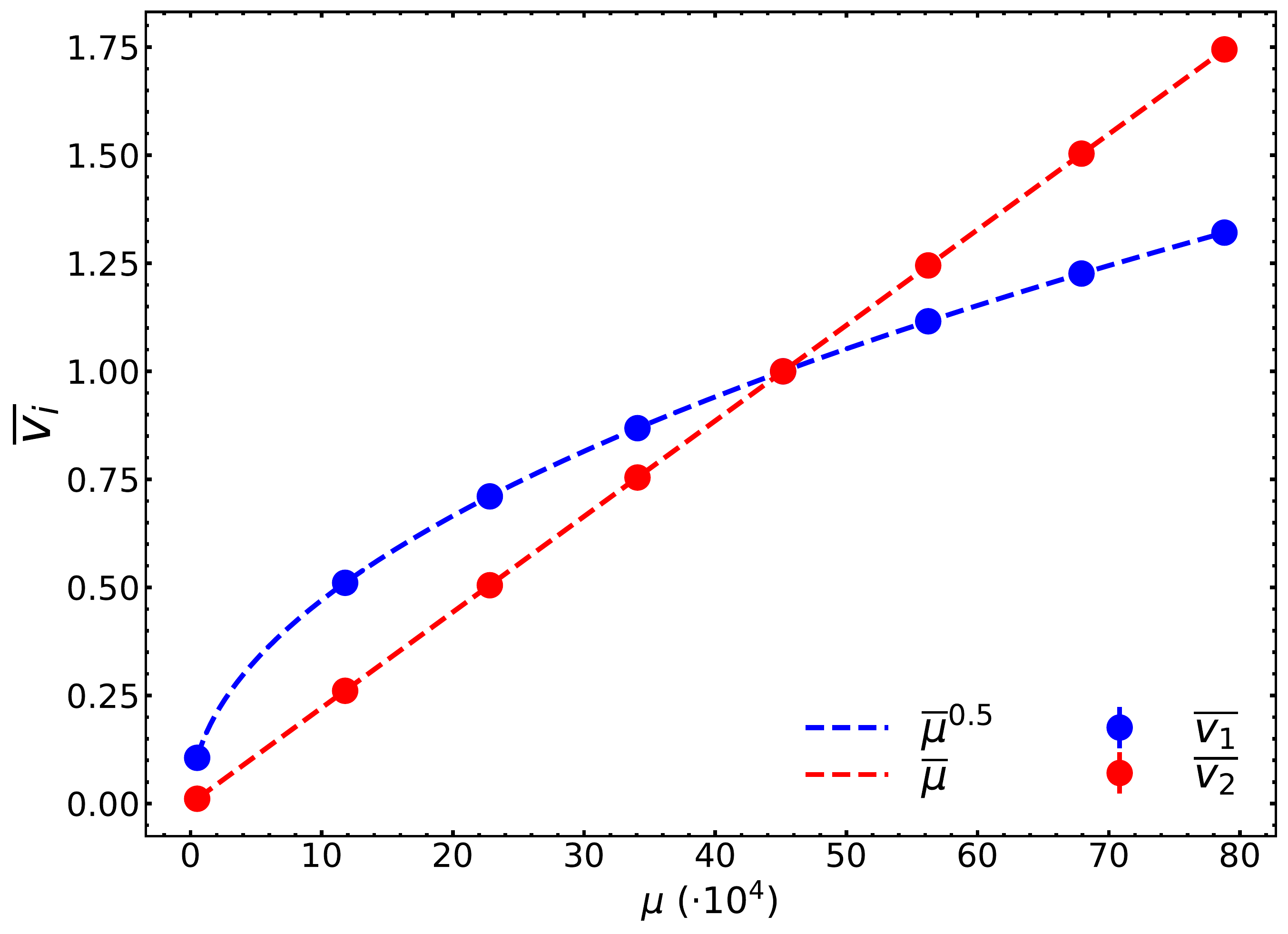} \caption{Dependence on $\mu$ of the normalised MFs $\overline{v_1}$ (blue dots) and $\overline{v_2}$ (red dots), defined in the text. We can see in dashed lines the curves for $\mu^{1/2}$ (blue) and $\mu$ (red), corresponding to the expected scaling relations.}
\label{fig:MFs_mus}
\end{figure}

The theoretical expectations of the MFs in \Cref{s:sotheo} assume a particular dependence of $\mu$ for the Lipschitz--Killing computed with different metrics (see the last paragraph in \Cref{sec:GKF}). In this section, we show that this scaling is strongly supported by simulations.

The assumed $\mu$ dependence has a direct impact on the predictions for $v_1$ and $v_2$. If the assumption was wrong, these MFs would present a different scaling with the parameter $\mu$. We test for deviations by using the MFs computed on monochromatic maps with a wide range of values of $\mu$.

In order to assess the dependence of $v_1$ and $v_2$ on $\mu$, we consider $\overline{v_i}=\Big\langle \frac{v_i}{v_i(\ell_m=950)}\Big\rangle$, where $\langle \cdot \rangle$ represents the average over the thresholds between $u=-3$ and $u=3$. By considering this quantity, we remove the dependence on threshold and constant factors, leaving only the dependence on $\mu$.

In \Cref{fig:MFs_mus} we show the trend of $\overline{v_1}$ and $\overline{v_2}$ for different values of $\mu$. We observe that they are perfectly compatible with the scaling assumed in the theory: $\mu^{1/2}$ and $\mu$, respectively.

Additionally, we have verified with the same procedure that $v_0$ and $v_3$ scale as $\mu^0$ and $\mu$, respectively, as predicted by the theory. All scaling relations remain unchanged by considering different thresholds and non--monochromatic maps, as expected.

\subsection{Gaussian CMB simulations}
\label{ss:gau_res}
In this section we verify that the theoretical predictions for the MFs (see \Cref{s:theory}) agree with the results computed on CMB Gaussian isotropic maps with a realistic angular power spectrum. We use the $300$ simulations introduced in \Cref{ss:simsgau}, which are generated with the Planck best--fit polarisation angular power spectra; we use the corresponding value of $\mu$ for the theoretical predictions. We compute the MFs (see \Cref{s:soimpl}) at thresholds between $u=-4$ and $u=4$, with a spacing of $\Delta u=0.2$. We use the values of the normalisation factors $K_1$ and $K_2$ obtained in monochromatic maps, \textit{i.e.}, \cref{eq:ks}. This means that there are no free parameters in the analysis of this section.

\begin{figure*}
	\centering
	\includegraphics[width=0.47\textwidth]{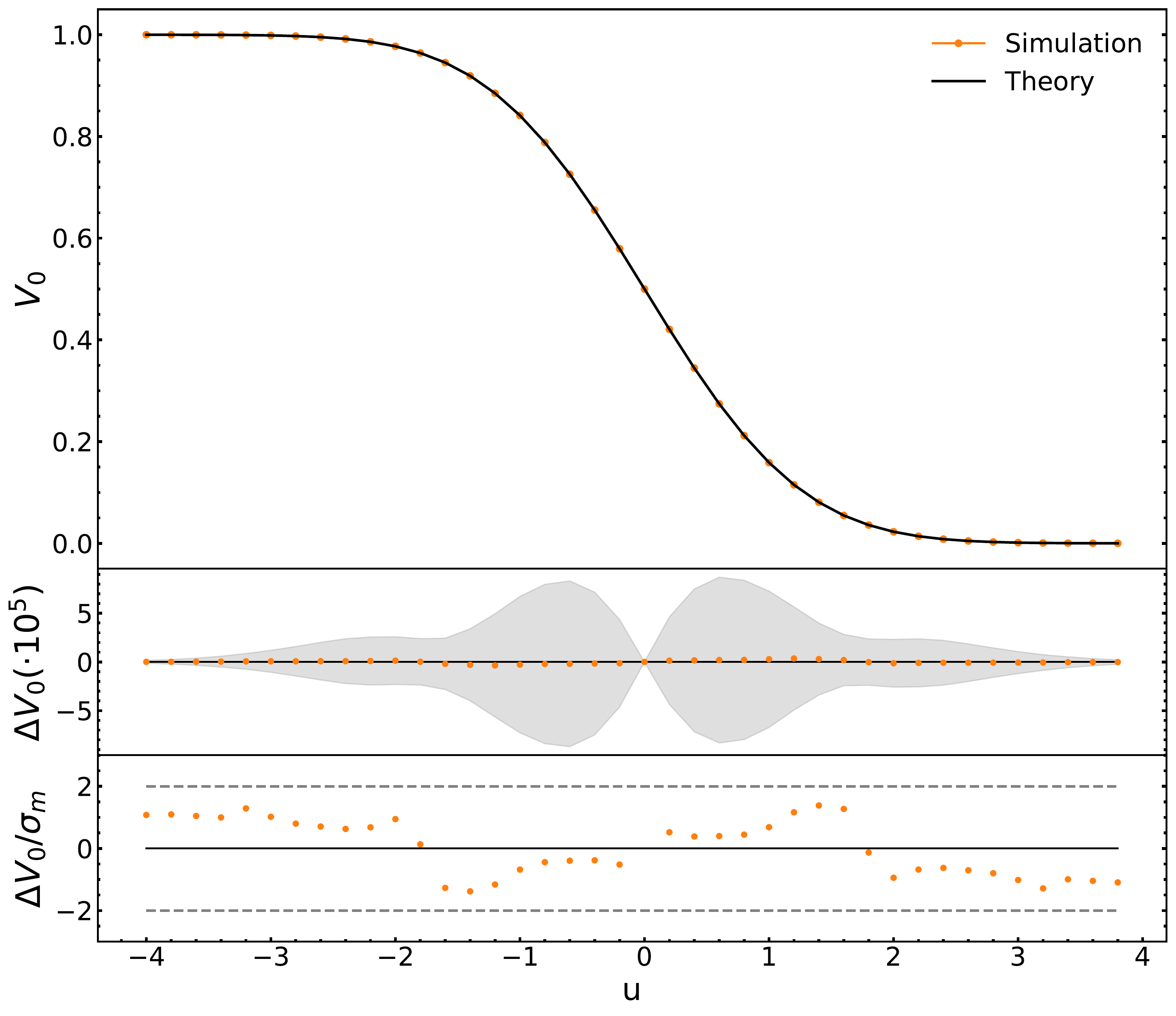}
    \hspace{0.4 cm}
	\includegraphics[width=0.48\textwidth]{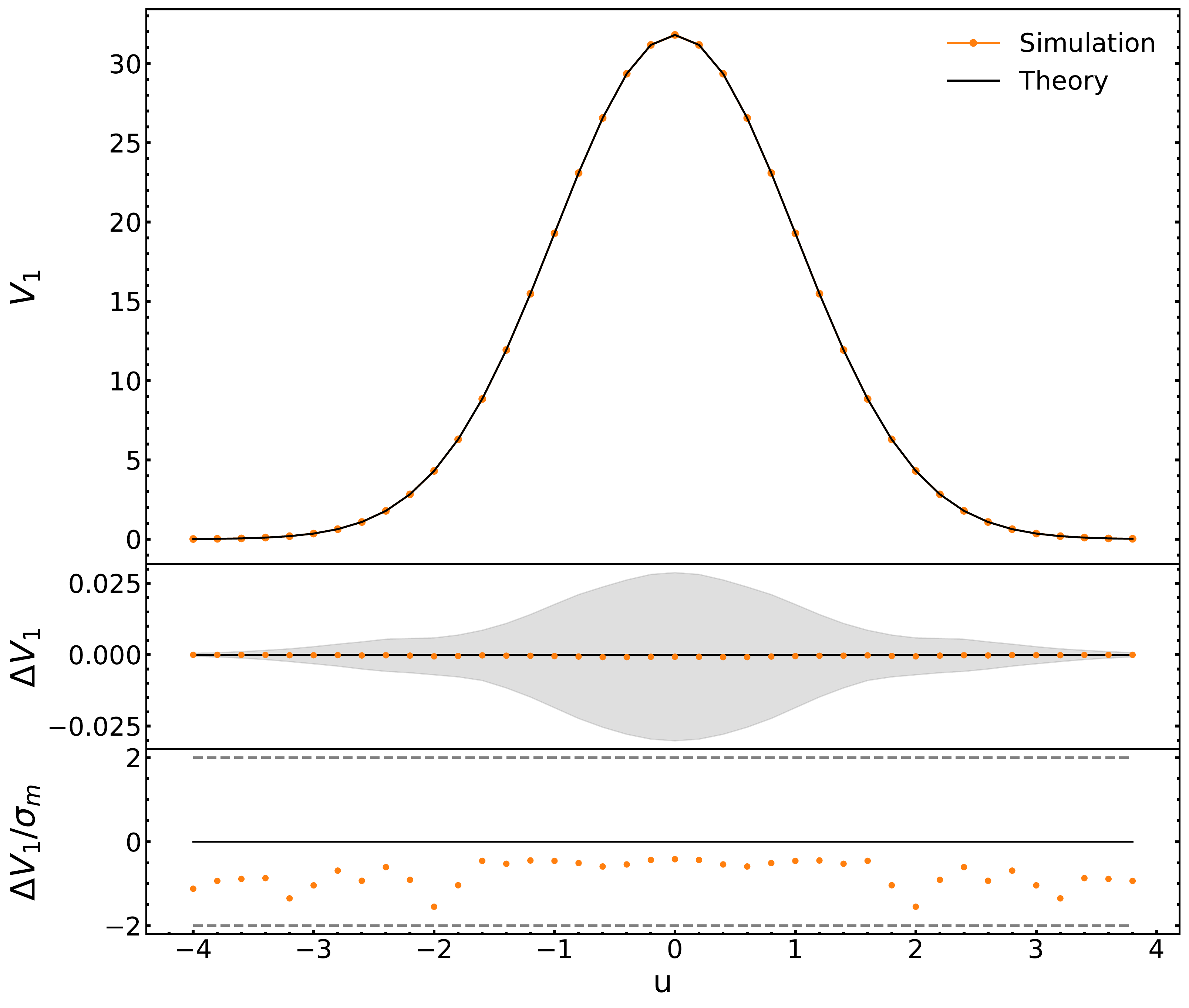} \\
	\hspace{-0.2 cm}\includegraphics[width=0.48\textwidth]{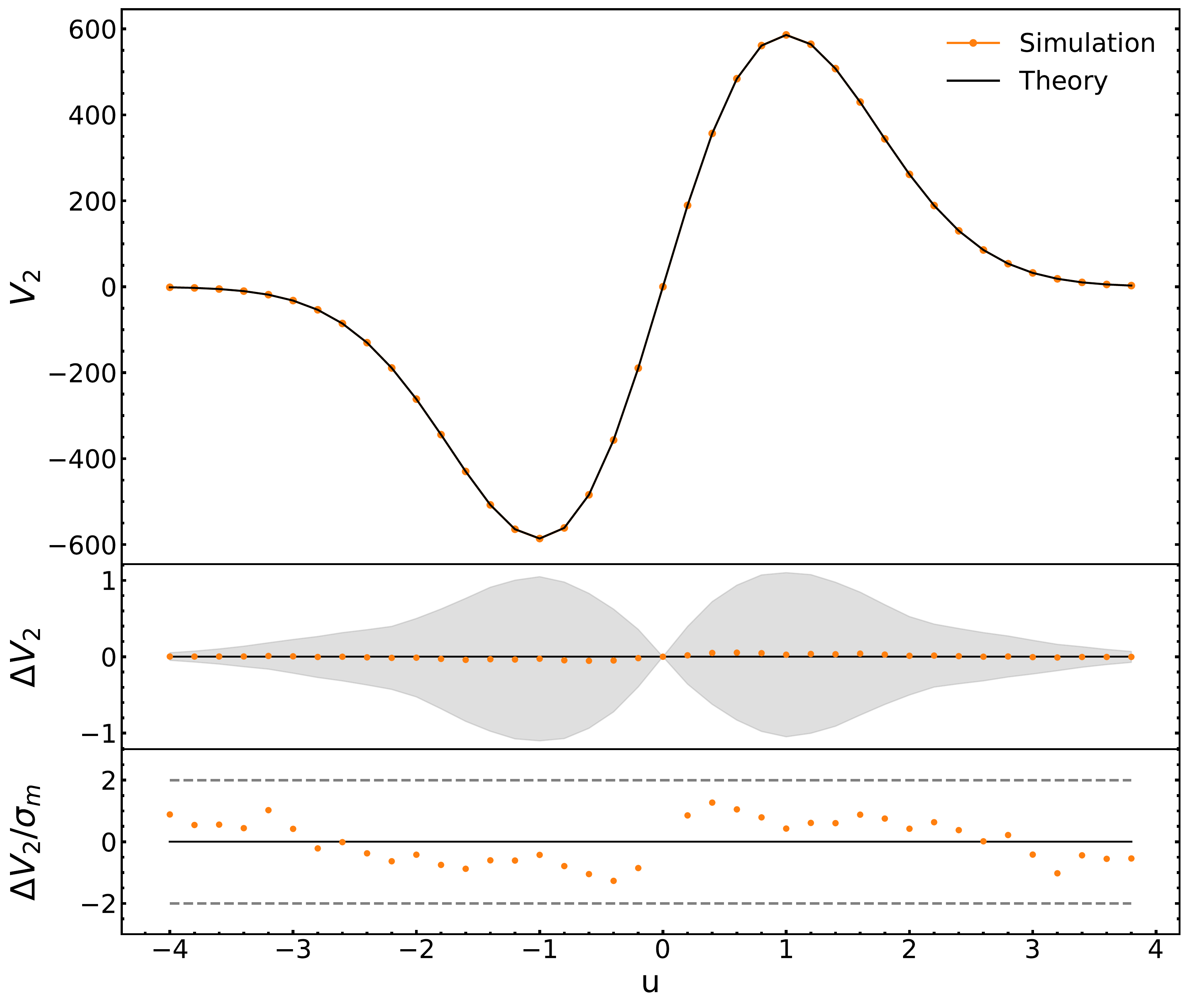}
    \hspace{0.45 cm}
	\includegraphics[width=0.477\textwidth]{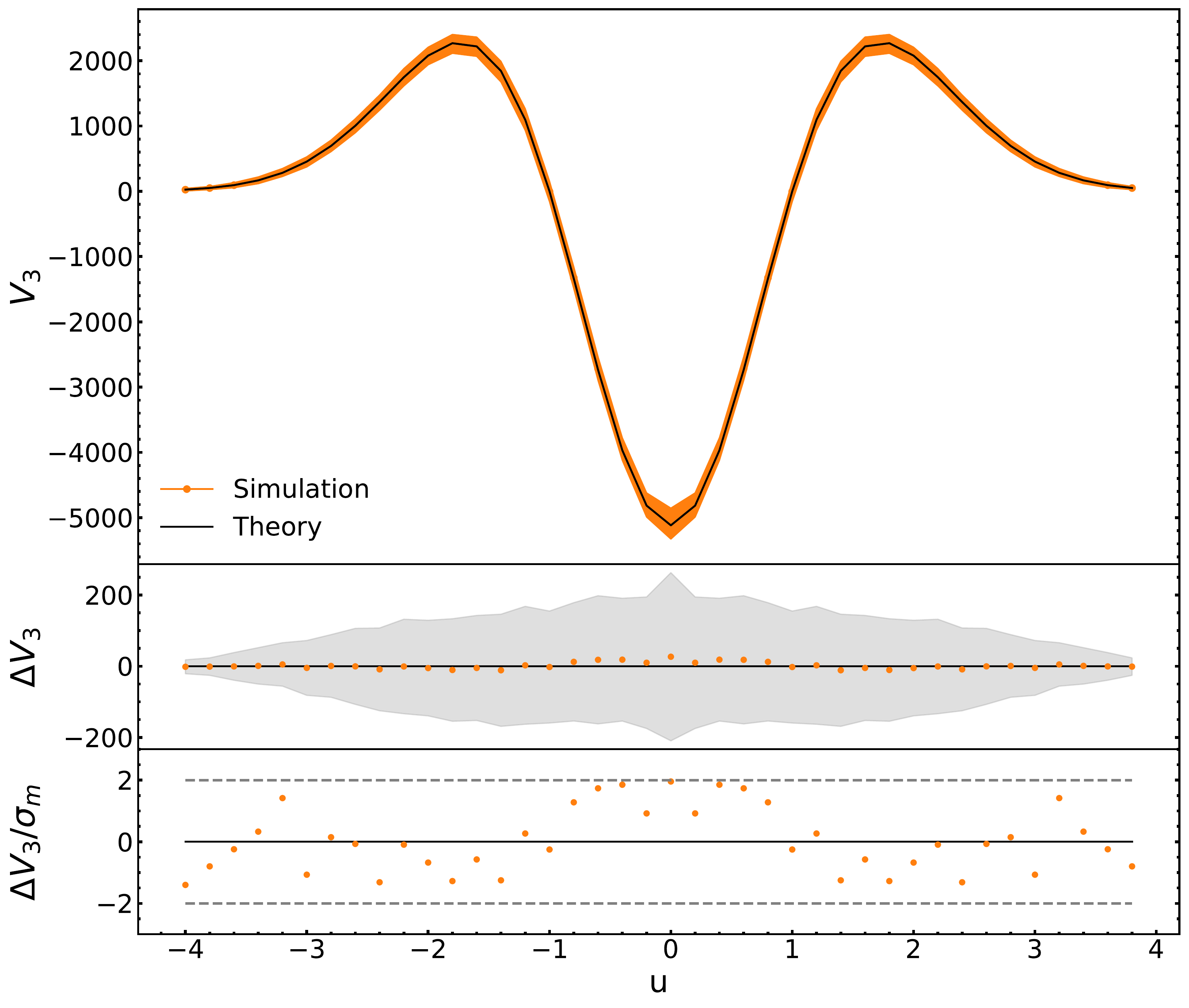}
	\caption{MFs for CMB polarisation simulations generated with the Planck best--fit angular power spectrum. Top left: $v_0$. Top right: $v_1$. Bottom left: $v_2$. Bottom right: $v_3$. In each panel we show, from top to bottom: the average value in simulations (orange)  and theoretical predictions (black), the difference between them compared with the standard deviation, $\sigma$, and the standard deviation of the mean, $\sigma_{m}$ (see text).}
\label{fig:MFs_cmb_gauss}
\end{figure*}
\Cref{fig:MFs_cmb_gauss} presents the comparison between the MFs computed on these maps and the corresponding theoretical predictions. We find full compatibility between theory and simulations for all MFs. In all cases, the residuals are well within the $1\sigma$ region, where $\sigma$ is the dispersion of the MFs among the $300$ simulations. 

Additionally, to explore any possible systematic deviation, we compare the average residual between simulations and theoretical expectations. This is done considering the standard deviation of the mean ($\sigma_{m} = \frac{\sigma_{\textrm{sims}}}{\sqrt{300}}$). This comparison is shown in the bottom sections of each panel in \Cref{fig:MFs_cmb_gauss}, in terms of $\Delta v_i/\sigma_m$, \textit{i.e.}, deviations over $\sigma_{m}$. It can be seen that no point deviates more than $2\sigma_{m}$, and there are no significant systematic residuals.

The four MFs are all perfectly compatible with the theory, in the case of both individual simulations and of average behaviour. We verify that the results hold for arbitrary angular power spectra, as long as there is no significant power in the scales corresponding to the pixel size.

We also note the low statistical variation of these curves: the relative uncertainty of every point is below $1$ part in $1 000$ for $v_{0}$, below $1\%$ for $v_1$ and $v_2$ and below $10\%$ for $v_3$. These values of the statistical standard deviation depend on the angular power spectrum of the studied map: we observe that the variance decreases when increasing the maximum multipole considered, $\ell_{max}$ (\textit{i.e.}, when increasing the resolution of the maps). This is also known to be the case for temperature maps, as quantitatively studied by \cite{fantaye2015}.

These results provide a double validation: on the one hand, they verify the mathematical theory used to predict the expected values of the MFs of $f$; on the other hand, they validate our implementation to compute these statistical quantities on polarisation data.

We have verified that masking the maps has a negligible effect on these results beyond slightly increasing the noise due to the smaller sky fraction. This robustness is expected because, unlike the angular power spectrum, MFs are purely local quantities. Thus, the effect of masking is not propagated.

\section{Conclusions}
\label{s:concl}
The study of the angular power spectrum of CMB anisotropies has led to the tightest constraints on the cosmological parameters \citep{2020A&A...641A...6P}. However, this statistical tool is not sensitive to important information which could be present at some level in CMB data, such as non-Gaussianities or deviations from statistical isotropy. The detection of such effects would reveal physical information about the Early Universe. 

In the past, MFs have been applied mainly to CMB temperature anisotropies and, more generally, to scalar fields in order to search for such signatures in Cosmological data \citep{schmalzing1998,planck2019vii,2022arXiv221107562C}.

In this work, we have extended the MFs formalism to the polarisation field by lifting it to the higher-dimensional manifold $SO(3)$, thus considering the full information embedded in this observable. In this framework, the complex part of the field is just a translation of the real one. Therefore, without loss of generality, we take only the real part, $f(\phi, \theta, \zeta) : SO(3) \rightarrow \mathbb{R}$ as:
\begin{equation}
	f(\phi, \theta, \zeta) = Q(\phi, \theta) \cos(2\zeta) - U(\phi, \theta)\sin(2\zeta),
\end{equation}
which is a scalar field that can be interpreted as the linear polarisation observed at the point $(\phi, \theta)$ on the sky in the polarisation direction $\zeta$.

To summarise:
\begin{itemize}
    \item We introduce a formalism to predict the values of MFs for $f$ in the case of Gaussian Stokes parameters through the use of the Gaussian Kinematic Formula for anisotropic fields (see Sects. \ref{sec:GKF} and \ref{s:theory}).
    \item We implement a code to compute MFs of $f$ from input Q and U maps pixelised with the HEALPix convention and to be released in the already publicly available Python package \href{https://github.com/javicarron/pynkowski}{\texttt{Pynkowski}}.
    \item We find that theoretical predictions are compatible with the computations on Gaussian CMB maps generated with the Planck best-fit angular power spectra.
\end{itemize}

The application on CMB polarisation data of statistical tools beyond the angular power spectrum (such as the MFs formalism introduced in this paper) will grow in importance with upcoming much more sensitive experiments from both ground and space, \textit{e.g.}, ACT \citep{2020JCAP...12..047A}, SPT \citep{2020PhRvD.101l2003S}, Simons Observatory \citep{2019JCAP...02..056A}, 
LiteBIRD \citep{2014JLTP..176..733M}, PICO \citep{2019arXiv190210541H}. 

We are currently applying this formalism to the CMB polarisation data from Planck and to simulated future experiments. The corresponding results will be presented in an upcoming paper.

Furthermore, this kind of analysis can help to blindly detect the presence of non-Gaussian contaminants in the maps like residual foregrounds contamination. It can also be used to characterise the morphology of Galactic emission, as recently done in \citep{2021rahman,martire2023} for synchrotron radiation.

The application proposed in this paper is versatile and can be performed on other spin--2 quantities. Moreover, in this work we have developed the predictions for Gaussian fields, but this framework can be extended to the weakly non--Gaussian case by following the approach adopted in \citep{matsubara2003} for fields defined on Euclidean spaces.

\section*{Acknowledgements}
The authors thank Michele Stecconi and Maurizia Rossi for insightful discussions. MM and NV acknowledge support by ASI/COSMOS grant n. 2016-24-H.0 and ASI/LiteBIRD grant n. 2020-9-HH.0. Part of this work was also supported by the InDark INFN project. DM acknowledges support from the MIUR Excellence Project awarded to the Department of Mathematics, Università di Roma Tor Vergata, CUP E83C18000100006. DM is also grateful to the Department of Excellence Programme MatModTov for support. This paper is supported by the Fondazione ICSC, Spoke 3 Astrophysics and Cosmos Observations, National Recovery and Resilience Plan (Piano Nazionale di Ripresa e Resilienza, PNRR) Project ID CN\_00000013 ``Italian Research Center on High-Performance Computing, Big Data and Quantum Computing'' funded by ``MUR Missione 4 Componente 2 Investimento 1.4: Potenziamento strutture di ricerca e creazione di campioni nazionali di R\&S (M4C2-19 ) - Next Generation EU (NGEU)''.

\bibliographystyle{JHEP}
\bibliography{biblio}

\newpage
\onecolumn
\appendix

\section{Polarization as a random field on $SO(3)$}
\label{ap:f}
In \Cref{s:lift}, we argued that the CMB polarisation spin field can be lifted to a complex function of $SO(3)$ and that its real and the imaginary parts are just translations of each other. Therefore, it suffices to consider only the real part, which corresponds to the scalar function $f$ introduced in \cref{eq:fdef}. In this appendix we will prove this statement. We note that this argument generalises to spherical fields of arbitrary spin.

We start by recalling the usual spherical harmonic decomposition of a spin--$s$ field:
\[
Q(\theta ,\phi )+iU(\theta ,\phi )=\sum_{\ell m}a_{\ell m;s} \: {}_{s}Y_{\ell m}(\theta ,\phi )\text{ ,} 
\]
with $_sY_{\ell m}(\theta ,\phi )$ the spin--weighted spherical harmonics. We can rewrite this decomposition by using the Wigner $D$--matrix $D_{\ell,m;s}$:
\[
Q(\theta ,\phi )+iU(\theta ,\phi )=\sum_{\ell m}a_{\ell m;s}D_{\ell,m;s}(\phi ,\theta ,0) \text{ .} 
\]%

The generalisation of this expression allows for the introduction of a three--dimensional complex field $\mathcal{P}$:
\[
\mathcal{P} (\phi ,\theta ,\zeta )= \mathcal{Q}(\phi ,\theta ,\zeta )+i\mathcal{U}(\phi ,\theta ,\zeta
)=\sum_{\ell m}a_{\ell m;s}D_{\ell ,m;s}(\phi ,\theta ,\zeta )\text{ ,} 
\]%
where $\mathcal{Q}$ and $\mathcal{U}$ are the Stokes parameters $Q$ and $U$ when the local reference frame is rotated by an angle $\zeta$: 
\begin{eqnarray*}
\mathcal{Q}(\phi ,\theta ,\zeta ) &=&Q(\theta ,\phi )\cos (s\zeta) -U(\theta
,\phi )\sin (s\zeta) \text{ ,} \\
\mathcal{U}(\phi ,\theta ,\zeta ) &=&U(\theta ,\phi )\cos (s\zeta) +Q(\theta
,\phi )\sin (s\zeta) \text{ .}
\end{eqnarray*}

The dependence of the Wigner $D$--matrix on its variables can be explicitly expressed by using the Wigner's $d$ function:
\[D_{\ell ,m;s}(\phi ,\theta ,\zeta ) = \exp (-im\phi )d_{\ell ,m;s}(\theta )\exp
(is\zeta)\]

We can see now that the following symmetry holds for the function $\mathcal{P}$:
\begin{eqnarray*}
\mathcal{P}(\phi ,\theta ,\zeta +\frac{\pi }{2s}) &=&\sum_{\ell m}a_{\ell
m;s}D_{\ell ,m;s}(\phi ,\theta ,\zeta ) \\
&=&\sum_{\ell m}a_{\ell m;s}\exp (-im\phi )d_{\ell ,m;s}(\theta )\exp
(is(\zeta +\frac{\pi }{2s})) \\
&=&i\sum_{\ell m}a_{\ell m;s}\exp (-im\phi )d_{\ell ,m;s}(\theta )\exp
(is\zeta ) \\
%&=& i \left( \mathcal{Q}(\phi ,\theta ,\zeta )+i\mathcal{U}(\phi ,\theta ,\zeta ) 
&=& i \: \mathcal{P}(\phi ,\theta ,\zeta) \text{ .}
\end{eqnarray*}%

Likewise:
\[
\mathcal{P}(\phi ,\theta ,\zeta -\frac{\pi }{2s})= -i \: \mathcal{P}(\phi ,\theta ,\zeta)  \text{ .}
\]%

In words, this means that a rotation of $\frac{\pi}{2s}$ of the local reference frame is equal to multiplying the function by a factor $i$ or, equivalently, swapping the real and imaginary components of $\mathcal{P}$: $\mathcal{Q}$ and $\mathcal{U}$. These components are therefore linked by the following expression:

\[
\mathcal{Q}(\phi ,\theta ,\zeta +\frac{\pi }{2s}) = Re\left[ \mathcal{P}(\phi ,\theta ,\zeta +\frac{\pi }{2s})\right] = -Im\left[ \mathcal{P}(\phi ,\theta ,\zeta)\right] = -\mathcal{U}(\phi ,\theta ,\zeta)
\]

Consequently, both the real and imaginary part of the lifted polarisation field $\mathcal{P}$ convey the same information, so it suffices to study one of them. We choose to study the real part $\mathcal{Q}$, which we call $f$ in the main text.

\section{Metric and derivatives in $SO(3)$}
\label{ap:SOhess}
In this paper, the introduced MFs formalism is based on the definition of the field $f(\phi, \theta, \zeta) = Q(\phi, \theta) \cos(2\zeta) - U(\phi, \theta)\sin(2\zeta)$, with domain in $SO(3)$. The geometry of this manifold is not trivial. In particular, we were unable to find in the literature the expressions for an orthonormal basis and for the Hessian of a function defined on this manifold. We performed these computations and we report them in this Appendix in case they are useful to the community.

$SO(3)$ can be seen as the group of rotations of the sphere, which can be parameterised by the Euler angles. These accept many different definitions, depending on the order of the axis selected to perform the rotations. To make the angles compatible with the usual coordinates on the sphere, we must select the definition in the $zyx$ convention. Then, a generic rotation of the sphere can be expressed as \citep[see, \textit{e.g.},][]{gelfand1963}:
\begin{equation*}
	R(\phi,\theta,\zeta) = R_x(\zeta) R_y(\theta) R_z(\phi),
\end{equation*}
where:
\begin{align*}
    R_x(\zeta)=&\left(\begin{matrix}1 & 0 & 0\\0 & \cos{\left(\zeta \right)} & - \sin{\left(\zeta \right)}\\0 & \sin{\left(\zeta \right)} & \cos{\left(\zeta \right)}\end{matrix}\right) ,\\
    R_y(\theta)=&\left(\begin{matrix}\cos{\left(\theta \right)} & 0 & \sin{\left(\theta \right)}\\0 & 1 & 0\\- \sin{\left(\theta \right)} & 0 & \cos{\left(\theta \right)}\end{matrix}\right) ,\\
    R_z(\phi)=&\left(\begin{matrix}\cos{\left(\phi \right)} & - \sin{\left(\phi \right)} & 0\\\sin{\left(\phi \right)} & \cos{\left(\phi \right)} & 0\\0 & 0 & 1\end{matrix}\right).
\end{align*}

It is then possible to compute the derivative of a random rotation with respect to the three parameters. Now, taking the Frobenius inner product between matrices (the sum of the element--wise multiplication of matrices), the metric of $SO(3)$ reads: 
\begin{equation*}
	g_{ij} = \langle \partial_i R , \partial_j R \rangle = 2 \left(\begin{matrix}1 & 0 & \sin{\left(\theta \right)}\\0 & 1 & 0\\ \sin{\left(\theta \right)} & 0 & 1\end{matrix}\right),
\end{equation*}
in the order $(\phi,\theta,\zeta)$. We can now determine the Christoffel symbols in this manifold:
\begin{align*}
	\Gamma^{\phi}_{\theta \zeta} = \Gamma^{\phi}_{\zeta \theta} = \Gamma^{\zeta}_{\phi \theta} = \Gamma^{\zeta}_{\theta \phi} &= \frac{1}{2\cos(\theta)}, \\
	\Gamma^{\phi}_{\theta \phi} = \Gamma^{\phi}_{\phi \theta} = \Gamma^{\zeta}_{\zeta \theta} = \Gamma^{\zeta}_{\theta \zeta} &= \frac{-\tan(\theta)}{2}, \\
	\Gamma^{\theta}_{\phi \zeta} = \Gamma^{\theta}_{\zeta \phi} &= \frac{-\cos(\theta)}{2},
\end{align*}
with the remaining ones equal to $0$. Using the Chistoffel symbols, the covariant derivatives are as follows:
\begin{equation*}
	\nabla_{\partial_a} \partial_b = \Gamma^c_{ab}\partial_c,
\end{equation*}
which, in our case, implies:
\begin{align*}
	\nabla_{\partial_\phi} \partial_\phi &= 0 ,\\
	\nabla_{\partial_\theta} \partial_\theta &= 0 ,\\
	\nabla_{\partial_\zeta} \partial_\zeta &= 0 ,\\
	\nabla_{\partial_\phi} \partial_\theta = \nabla_{\partial_\theta} \partial_\phi &= - \frac{\tan(\theta)}{2} \partial_\phi + \frac{1}{2\cos(\theta)} \partial_\zeta ,\\
	\nabla_{\partial_\zeta} \partial_\theta = \nabla_{\partial_\theta} \partial_\zeta &= \frac{1}{2\cos(\theta)} \partial_\zeta - \frac{\tan(\theta)}{2} \partial_\phi ,\\
	\nabla_{\partial_\phi} \partial_\zeta = \nabla_{\partial_\zeta} \partial_\phi &= - \frac{\cos(\theta)}{2} \partial_\theta.
\end{align*}

These covariant derivatives can then be employed to compute the components of the Hessian according to:

\begin{align*}
	&\hess_{ij}(f) = \nabla^2 f(e_{i},e_{j})= e_i e_j f - \nabla_{e_i} e_j f ,
\end{align*}

For the first part of each term, we obtain:

\begin{align*}
    e_1 e_1 = &\Bigg(\eone\Bigg)\\
     &\Bigg(\eone\Bigg) \\
    = &\frac{1}{4 \cos^{2}{\left(\theta \right)}} %
	\bigg[
	\left(1+\cos{\left(\theta \right)}\right) \frac{\partial^{2}}{\partial \phi^{2}} + \left(1-\cos{\left(\theta \right)}\right) \frac{\partial^{2}}{\partial \zeta^{2}} - 2 \sin{\left(\theta \right)} \frac{\partial^{2}}{\partial \zeta\partial \phi} 
	\bigg] ,\\
    e_2 e_2 =& \Bigg(\etwo\Bigg)\Bigg(\etwo\Bigg) = \frac{\partial^{2}}{2\partial \theta^{2}},\\
    e_3 e_3 =& \Bigg(\ethree\Bigg)\\
     &\Bigg(\ethree\Bigg) \\
    =& \frac{1}{4 \cos^{2}{\left(\theta \right)}} %
		\bigg[
		\left(1-\cos{\left(\theta \right)}\right) \frac{\partial^{2}}{\partial \phi^{2}} + \left(1+\cos{\left(\theta \right)}\right) \frac{\partial^{2}}{\partial \zeta^{2}} - 2 \sin{\left(\theta \right)} \frac{\partial^{2}}{\partial \zeta\partial \phi} 
		\bigg] ,\\
    e_1 e_2 = e_2 e_1 =& \Bigg(\eone\Bigg)\Bigg(\etwo\Bigg) \\
    =& \frac{1}{4 \cos{\left(\theta \right)}} \Bigg[ \left(\sqrt{1 - \sin{\left(\theta \right)}} - \sqrt{\sin{\left(\theta \right)} + 1}\right) \frac{\partial^{2}}{\partial \theta\partial \zeta} + \left(\sqrt{1 - \sin{\left(\theta \right)}} + \sqrt{\sin{\left(\theta \right)} + 1}\right) \frac{\partial^{2}}{\partial \theta\partial \phi} \Bigg] ,
\end{align*}
\begin{align*}
    e_1 e_3 = e_3 e_1 =& \Bigg(\eone\Bigg)\\
    &\Bigg(\ethree\Bigg) \\
    =& \frac{1}{4 \cos^2{\left(\theta \right)}} \Bigg[ - \sin{\left(\theta \right)} \frac{\partial^{2}}{\partial \phi^2} - \sin{\left(\theta \right)} \frac{\partial^{2}}{\partial \theta^2} + 2 \frac{\partial^{2}}{\partial \phi\partial \zeta}  \Bigg] ,\\
    e_2 e_3 = e_2 e_3 =& \Bigg(\etwo\Bigg)\Bigg(\ethree\Bigg) \\
    =& \frac{1}{4 \cos{\left(\theta \right)}} \Bigg[ \left(\sqrt{1 - \sin{\left(\theta \right)}} - \sqrt{\sin{\left(\theta \right)} + 1}\right) \frac{\partial^{2}}{\partial \theta\partial \zeta} + \left(\sqrt{1 - \sin{\left(\theta \right)}} + \sqrt{\sin{\left(\theta \right)} + 1}\right) \frac{\partial^{2}}{\partial \theta\partial \phi} \Bigg],
\end{align*}

while the second part of each term is given by:
\begin{align*}
    \nabla_{e_1} e_1 = &\nabla_{\eone}\Bigg(\eonephi + \\
    &+ \eonezeta\Bigg) = \frac{\sin{\left(2 \theta \right)} }{8 \cos^{2}\theta} \frac{\partial}{\partial \theta},\\
    \nabla_{e_2} e_2 =& \nabla_{\etwo}\Bigg(\etwo\Bigg) = 0,\\
    \nabla_{e_3} e_3 = &\nabla_{\ethree}\Bigg(\ethreephi + \\
    &+ \ethreezeta\Bigg)= \frac{\sin{\left(2 \theta \right)} }{8 \cos^{2}\theta} \frac{\partial}{\partial \theta} ,\\
  %    \end{align*}
  % \begin{align*}
    \nabla_{e_1} e_2 = \nabla_{e_2} e_1 =& \nabla_{\eone}\Bigg(\etwo\Bigg) \\
    =& \frac{1}{8 \cos^{2}{\left(\theta \right)}}
		\bigg[  \big(P(\theta)\sin{\left(\theta \right)} - M(\theta) \big) \frac{\partial}{\partial \phi}   + \big(M(\theta)\sin{\left(\theta \right)} - P(\theta) \big) \frac{\partial}{\partial \zeta}  \bigg] ,\\
    \nabla_{e_1} e_3 = \nabla_{e_3} e_1=& \nabla_{\eone}\Bigg(\ethreephi + \\
    &+ \ethreezeta\Bigg) = - \frac{1}{4 \cos{\left(\theta \right)}} \frac{\partial}{\partial \theta} ,\\
    \nabla_{e_2} e_3 = \nabla_{e_3} e_2 = &\nabla_{\Big(\ethree\Big)}\Bigg(\etwo\Bigg) \\
    =& \frac{1}{8 \cos^{2}{\left(\theta \right)}}
		\bigg[  \big(M(\theta)\sin{\left(\theta \right)} - P(\theta) \big) \frac{\partial}{\partial \phi}   + \big(P(\theta)\sin{\left(\theta \right)} - M(\theta) \big) \frac{\partial}{\partial \zeta}  \bigg] ,
\end{align*}

where:
%\begin{subequations}
\begin{align*}
        P(\theta) \coloneqq& \sqrt{1 - \sin{\left(\theta \right)}} + \sqrt{\sin{\left(\theta \right)} + 1} ,\\
		M(\theta) \coloneqq& \sqrt{1 - \sin{\left(\theta \right)}} - \sqrt{\sin{\left(\theta \right)} + 1}.
\end{align*} %\label{eq:sohess}
%\end{subequations}

Therefore, the components of the Hessian are the following:
%\begin{subequations}
\begin{align*}
	\hess_{11} =& \frac{1}{4 \cos^{2}{\left(\theta \right)}} %
		\bigg[
		\left(1+\cos{\left(\theta \right)}\right) \frac{\partial^{2}}{\partial \phi^{2}} + \left(1-\cos{\left(\theta \right)}\right) \frac{\partial^{2}}{\partial \zeta^{2}} - 2 \sin{\left(\theta \right)} \frac{\partial^{2}}{\partial \zeta\partial \phi} 
		- \frac{\sin{\left(2 \theta \right)} }{2} \frac{\partial}{\partial \theta}
		\bigg] ,\\
		\hess_{22} = &\frac{\partial^{2}}{2\partial \theta^{2}},\\
		\hess_{33} =& \frac{1}{4 \cos^{2}{\left(\theta \right)}} %
		\bigg[
		\left(1-\cos{\left(\theta \right)}\right) \frac{\partial^{2}}{\partial \phi^{2}}   + \left(1+\cos{\left(\theta \right)}\right) \frac{\partial^{2}}{\partial \zeta^{2}} - 2 \sin{\left(\theta \right)} \frac{\partial^{2}}{\partial \zeta\partial \phi} -\frac{\sin{\left(2 \theta \right)} }{2} \frac{\partial}{\partial \theta}
		\bigg] ,\\
		\hess_{12} = \hess_{21}=& \frac{1}{8 \cos^{2}{\left(\theta \right)}} 
		\bigg[ 
		2\cos{\left(\theta \right)} \left(M(\theta) \frac{\partial^{2}}{\partial \theta\partial \zeta} 
			+ P(\theta) \frac{\partial^{2}}{\partial \theta\partial \phi}\right) + \left( P(\theta) \sin{\left(\theta \right)} - M(\theta) \right) \frac{\partial}{\partial \phi} + \\ 
		&+ \left( M(\theta) \sin{\left(\theta \right)} - P(\theta) \right) \frac{\partial}{\partial \zeta}
		\bigg] ,\\
		\hess_{13} = \hess_{31}= &\frac{1}{4 \cos^{2}{\left(\theta \right)}}
		\bigg[ -
		\sin{\left(\theta \right)} \frac{\partial^{2}}{\partial \phi^{2}} 
		- \sin{\left(\theta \right)} \frac{\partial^{2}}{\partial \zeta^{2}} + 2 \frac{\partial^{2}}{\partial \zeta\partial \phi}
		+ \cos{\left(\theta \right)} \frac{\partial}{\partial \theta} 
		\bigg] ,\\
		\hess_{23} = \hess_{32} =&  \frac{1}{8 \cos^{2}{\left(\theta \right)}}
		\bigg[ 
		2 \cos{\left(\theta \right)} \left(M(\theta) \frac{\partial^{2}}{\partial \theta\partial \phi} + P(\theta) \frac{\partial^{2}}{\partial \theta\partial \zeta} \right)  + \big(M(\theta)\sin{\left(\theta \right)} - P(\theta) \big) \frac{\partial}{\partial \phi}   + \\
  &+ \big(P(\theta)\sin{\left(\theta \right)} - M(\theta) \big) \frac{\partial}{\partial \zeta}  
		\bigg] .
\end{align*}

\end{document}